\documentclass[3p,review]{elsarticle}
\usepackage{amsmath}
\usepackage{amssymb}
\usepackage{multirow}
\usepackage{graphicx}
\usepackage{siunitx}
\usepackage{placeins}
\usepackage{dcolumn}
\usepackage{bm}
\usepackage{txfonts}
\usepackage[dvipsnames]{xcolor}
\usepackage[english]{babel}
\usepackage[normalem]{ulem}
\usepackage{booktabs}
\usepackage{soul}
\setulcolor{red}
\usepackage[linkcolor=blue,urlcolor=blue,citecolor=blue,colorlinks=true]{hyperref}

\journal{Applied Surface Science}
\begin{document}

\title{Revealing the (111) surface electronic structure of epitaxially grown Na$_2$KSb photocathode}

\author[isp,sync]{N. Yu. Solovova\corref{cor1}}
\ead{solovovanadezda822@gmail.com}

\author[isp,sync]{V. A. Golyashov}

\author[ispm]{S. V. Eremeev}

\author[ioffe]{S. Yu. Priobrazhenskii}

\author[ioffe]{S. P. Lebedev}

\author[ioffe]{A. A. Lebedev}

\author[isp,ekran]{V. S. Rusetsky}

\author[isp,sync]{O. E. Tereshchenko}
\ead{teresh@isp.nsc.ru}

\cortext[cor1]{Corresponding author}

\address[isp]{Rzhanov Institute of Semiconductor Physics, Siberian Branch, Russian Academy of Sciences, Novosibirsk, 630090, Russia}
\address[sync]{Synchrotron radiation facility SKIF, Boreskov Institute of Catalysis, Siberian Branch, Russian Academy of Sciences, Kol'tsovo, 630559, Russia}
\address[ispm]{Institute of Strength Physics and Materials Science, Siberian Branch, Russian Academy of Sciences, Tomsk, 634055, Russia}
\address[ioffe]{Ioffe Institute, 26 Politekhnicheskaya, 194021 St Petersburg, Russia}
\address[ekran]{JSC "EKRAN FEP", Novosibirsk 630060, Russia}

\date{\today}

\begin{abstract}
A recent study has established the Na$_2$KSb(Cs) photocathode as a highly efficient emitter of spin-polarized electrons. However, the electronic structure of alkali antimonides remains poorly understood. In this work, we report the first crystalline epitaxial growth of Na$_2$KSb films, achieved via chemical vapor deposition (CVD) on a graphene-coated SiC(0001) substrate. The high crystalline quality of these films enabled a direct investigation of the material's electronic structure using angle-resolved photoemission spectroscopy (ARPES). By comparing the experimental results with density functional theory (DFT) calculations, we have identified dispersive surface states originating from different terminations of the Na$_2$KSb(111) surface. Furthermore, we demonstrate that the crystalline order of the film is preserved following its activation via the deposition of Cs and Sb. This finding opens a pathway for investigating the electronic structure of multialkali Na$_2$KSb(Cs) photocathodes and for rationally improving their properties.

\end{abstract}

\begin{keyword}
 
Na$_2$KSb \sep multi-alkali antimonides \sep semiconductor photocathodes \sep electronic structure \sep surface states \sep ARPES
\end{keyword}

\maketitle


\section{\label{sec:level1}Introduction}
Photocathodes are essential components in various photon detectors, accelerators \cite{Electronsour, Miyoshi:2010mya}, electron microscopes \cite{n2002spin, Krysztof}, and other modern technologies. Several classes of materials are used for photocathodes, each with its own advantages and disadvantages \cite{schaber2023review, PhotocathodesfHB}. Therefore, the specific requirements for the photocathode determine the material of choice.

Alkali antimonides are widely used semiconductor photocathodes due to their high quantum efficiency (QE), low mean transverse energy, and lower sensitivity to residual gases in the vacuum chamber compared to the popular GaAs photocathode \cite{sommer2001brief, interface, cultrera_gulliford_bartnik_lee_bazarov_2016, rozhkov2025}. Until recently, spin-dependent photoemission upon absorption of circularly polarized light was known only for GaAs(Cs,O) photocathodes, making them attractive for spin-polarized electron injection applications. However, this effect has recently been demonstrated in a Na$_2$KSb(Cs) photocathode (also known as S20/S25) \cite{rusetsky2022new,PRL_spin_filter}, which was shown to exhibit even higher spin polarization than the well-known GaAs(Cs,O) photocathode.

As established in the literature, Na$_2$KSb has a face-centered cubic (FCC) lattice.\cite{scheer1959crystal} However, Na$_2$KSb-based photocathodes grown on conventional substrates such as glass or metals are believed to have a polycrystalline structure\cite{interface}, and currently no established technique exists for achieving crystalline ordered growth of any multi-alkali antimonide. The development of such epitaxial techniques is motivated by several key challenges. For instance, single-crystal photocathodes are critical for high-brightness photoinjectors, as the conservation of transverse momentum during photoemission from an ordered surface enables the generation of electron beams with very low mean transverse energy \cite{karkare2020ultracold}. Additionally, the electronic band structure of multi-alkali antimonides has so far been studied only by first-principles calculations \cite{ettema2000electronic, amador2021electronic} and indirect methods such as optical spectroscopy, which lacks momentum resolution. Consequently, the understanding of the spin-polarized electron emission mechanism in these materials still relies predominantly on calculated band structures. Moreover, surface states in multi-alkali antimonides remain poorly studied, even though they can significantly influence photoemission for both bandgap and subbandgap excitation \cite{K2CsSb_surf, wang2022first}. The surface electronic structure of Na$_2$KSb and Na$_2$KSb(Cs) has not been studied at all. Angle-resolved photoemission spectroscopy (ARPES) is a powerful experimental technique for investigating surface electronic structure, but it requires samples with a single-crystal surface \cite{sobota2021angle}. Furthermore, the development of epitaxial growth for multi-alkali antimonide photocathodes could enable more efficient spin-polarized electron sources. It is known, for example, that the spin polarization of electrons emitted from GaAs(Cs,O) photocathodes can be enhanced using strained-layer superlattices such as GaAs/GaAsP, where uniaxial strain induces energy splitting of the heavy- and light-hole bands \cite{biswas2023record, Highly_polarized}. Successful epitaxial growth of multi-alkali antimonides could open a pathway to similarly engineer the spin polarization of electrons emitted from Na$_2$KSb-based photocathodes.

To date, single-crystal growth has been demonstrated for the mono-alkali antimonides CsSb \cite{parzyck2023atomically} and Cs$_3$Sb \cite{parzyck2022single, pennington2025structural}. These studies noted that growth temperature and substrate material significantly impact the crystalline structure of the cathode. In particular, 3C-SiC(001) has been identified as a suitable substrate for the epitaxial growth of alkali antimonides. The single-crystal growth of multi-alkali antimonides remains almost unexplored. To our knowledge, only one study \cite{flintsynthesis} has reported the possibility of 3D island growth of Na$_2$KSb on Si(111) and Si$_3$N$_4$ substrates. 

Here, we demonstrate the first epitaxial growth of the multialkali antimonide Na$_2$KSb on graphene-coated SiC(0001) substrate and study Na$_2$KSb(111) surface electronic structure using ARPES and density functional theory (DFT) calculations. As in studies of CsSb and Cs$_3$Sb, Molecular Beam Epitaxy (MBE) can be used for Na$_2$KSb growth\cite{balanyuk1988synthesis, massegu2008s20}. In this work, however, we applied a standard chemical vapor deposition (CVD) process \cite{erjavec1994alkali}.

\section{Experimental and computational details}

\subsection{\label{sec:level2}UHV equipment and measurements}
\begin{figure*}[htbp]
\includegraphics[width=0.99\textwidth]{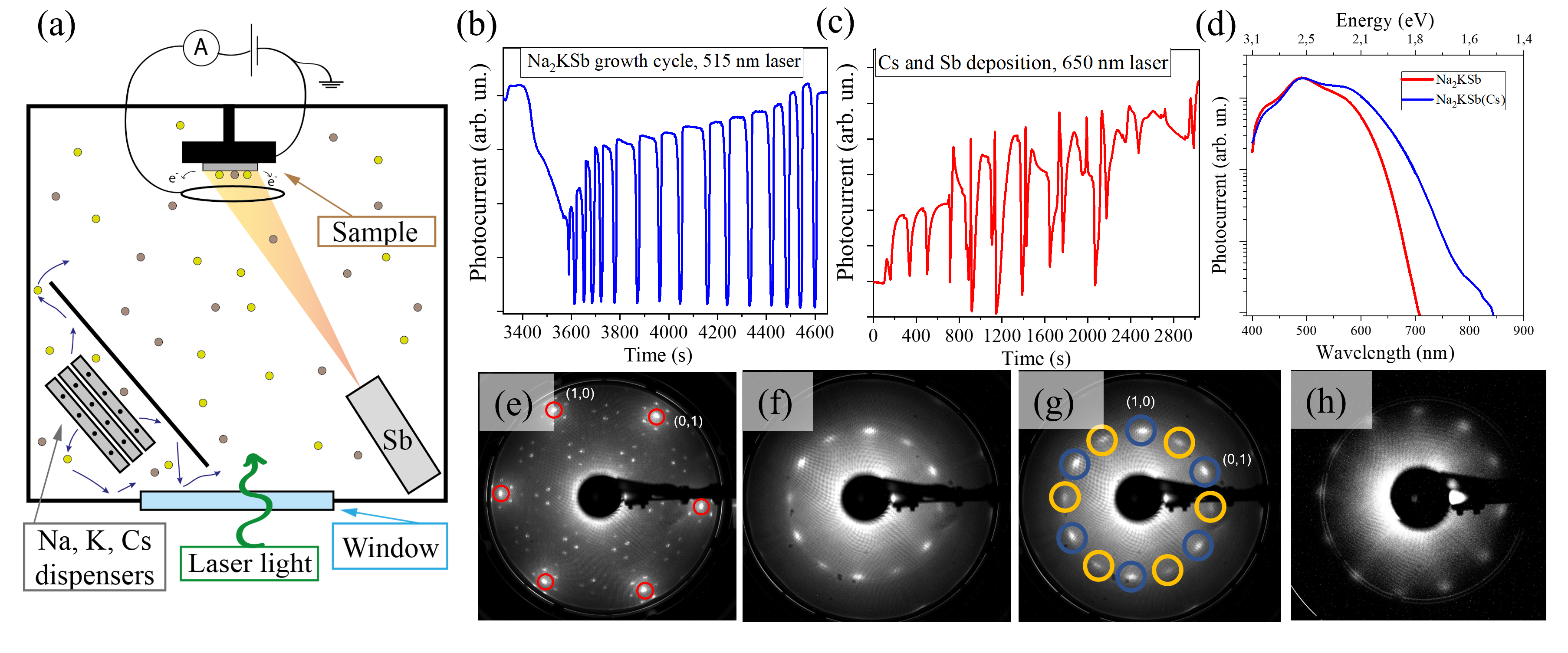}
\caption{\label{figpic4}(a) Scheme of the Na$_2$KSb(Cs) growth process; (b) Change in photocurrent during one growth cycle of Na$_2$KSb; (c) Change in photocurrent during activation of the Na$_2$KSb surface; (d) Spectral dependence of the photocurrent measured on Na$_2$KSb and Na$_2$KSb(Cs); LEED pattern observed on the surface (e) of graphene/SiC(0001) substrate, $E=80$ eV (f) Na$_2$KSb films $E=34$ eV; (g) the contribution of two different domains to the LEED pattern (f) is highlighted in blue and yellow; (h) LEED pattern observed after the Na$_2$KSb surface activation.}
\end{figure*}

Sample growth and all measurements were performed in an ultrahigh vacuum (UHV) at base pressures of $10^{-10}$ mbar and $10^{-11}$ mbar, respectively. Fig. \ref{figpic4}(a) shows a schematic diagram of the growth chamber, which is typical for the CVD synthesis of Na$_2$KSb \cite{bates1979basic}. The photocathode growth was carried out in an enclosed vessel inside the vacuum chamber. The substrate and vessel temperatures were maintained at 180–200 $^{\circ}$C during the Na$_2$KSb deposition and at 120–140 $^{\circ}$C during the activation process (Sb and Cs sputtering). Cs, Na, and K were evaporated from SAES-type dispensers \cite{SAESdisp} onto the vessel walls and subsequently re-evaporated onto the sample surface due to their high saturated vapor pressure. The dispensers were heated to activate the chemical reaction and initiate alkali metal evaporation prior to photocathode fabrication. Sb was deposited directly onto the sample surface from an effusion cell. Epitaxial graphene was produced by the SiC surface sublimation method \cite{ma14030590}. High-purity semi-insulating single-crystal 6H-SiC wafers were employed for the synthesis. A crystallographic orientation of the SiC growth face (0001) $\pm$ 0.25$^{\circ}$ (Si-face) for growth of monolayer graphene films was chosen. The growth surface of the SiC wafers was processed using chemical-mechanical polishing (CMP) technology. Epitaxial graphene was synthesized in a high frequency induction heating system furnished with graphite crucible: the SiC wafer was heated in a chamber filled with 6N Ar at the pressure of 750 $\pm$ 20 Torr up to 1730 $\pm$ 20 $^{\circ}$C at a rate about 100 $^{\circ}$C/min and then annealed for 5 min. The surface of the epitaxial graphene/6H-SiC(0001) substrates was cleaned by annealing at 500–600 $^{\circ}$C for 30 minutes in UHV. To monitor the photocathode growth process, the photocurrent was measured, as it provides a standard method for  controlling the evaporation of Na, K, Cs, and Sb \cite{galdi2019towards, erjavec1994alkali}.  Lasers with wavelengths of 515 nm and 650 nm were used during the growth of the Na$_2$KSb and (Cs,Sb) layers, respectively, to excite the photocurrent.

ARPES and X-ray photoelectron spectroscopy (XPS) measurements were conducted using a SPECS GmbH ProvenX-ARPES system equipped with an ASTRAIOS 190 electron energy analyzer and a 2D-CMOS electron detector. ARPES measurements were performed with nonmonochromated He I$\alpha$ light ($h\nu$ = 21.22 eV). XPS measurements were made using focused monochromatic AlK$\alpha$  radiation ($h\nu$ = 1486.7 eV, 0.5 mm X-ray spot at 180 W anode power). XPS spectra were recorded at normal emission and a constant pass energy of 30 eV with a total energy resolution of $<0.6$ eV.

\subsection{\label{sec:level2}DFT calculations}

Electronic structure calculations were carried out within the DFT as implemented in the VASP code \cite{vasp1,vasp2}, where the generalized gradient approximation (GGA-PBE) \cite{Perdew.prl1996} for the exchange-correlation potential was applied and the interaction between the ion cores and valence electrons was described by the projector augmented-wave method~\cite{Blochl.prb1994}. The alkali metals $p$ orbitals were treated as valence electrons.  Relativistic effects, including the spin-orbit interaction, were taken into account both during structure relaxation and band spectra calculations.  The crystal structure was fully optimized to find the equilibrium lattice parameters and atomic positions. At that, a conjugate-gradient algorithm was used. The DFT-D3 van der Waals (vdW) functional with Becke-Johnson damping \cite{Grimme2011} were applied for structure optimization. The alkaly-atom terminated Na$_2$KSb(111) surfaces were modeled using symmetrical slabs separated by the vacuum gap in the repeated-slab approach with a thickness of 45 atomic layers for K-terminated surface, 47 and 43 atomic layers for Na-K- and Na-Sb-terminations (hereinafter referred to as Na$_1$ and Na$_2$, respectively). The vacuum gap was chosen to be 15 {\AA} in all considered slabs. Six atomic layers on both surfaces of the slabs were allowed to relax while atoms in the internal layers were kept in their equilibrium bulk positions.   The $k$-point grids of 10$\times10\times$10 and 10$\times10\times$1 were used to sample the bulk and surface Brillouin zones, respectively. Similar to the Na$_2$KSb bulk calculations \cite{rusetsky2022new} for surface band structure calculations we use Slater-type DFT-1/2 self-energy correction method \cite{DFT12_1,DFT12_2} for construction of partially ionized antimony potential to obtain realistic bulk band gap.

The presented ball-and-stick atomic structures were visualized with {\sc vesta} \cite{VESTA}.

\section{\label{sec:level1}Results and discussion}

\subsection{\label{sec:level2}Na$_2$KSb epitaxial growth}

The growth of Na$_2$KSb and its subsequent activation were based on the standard photocathode S25 synthesis procedure described elsewhere \cite{erjavec1994alkali}. In this method, the main increase in film thickness is achieved through repeated growth cycles. Fig. \ref{figpic4}(b) thus shows the photocurrent changes within one cycle. Each cycle begins with the evaporation of Na, followed by the periodic flash-evaporation of Sb in K vapors. Each dip observed in the photocurrent curve corresponds to a flash-evaporation of Sb. A total of 6 cycles were performed, resulting in a final film thickness of approximately 40 nm. Once the required thickness of the Na$_2$KSb film was achieved, the cyclic process was terminated, and the sample was cooled to a temperature of 140 $^{\circ}$C. The activation process via Cs and Sb deposition was similar to that described in the literature \cite{erjavec1994alkali}, but was conducted at a lower temperature, decreasing from 140 $^{\circ}$C to 120 $^{\circ}$C during the process. The changes in photocurrent during the activation are shown in Fig. \ref{figpic4}(c). As seen in the Fig. \ref{figpic4}(c), the photocurrent under red laser illumination (650 nm) increases during this process. Figure \ref{figpic4}(d) shows the spectral dependence of photocurrent (normalized to the photon flux) of the Na$_2$KSb film before and after surface activation with Cs and Sb. A slower decay of the photocurrent at longer wavelengths is observed after activation. This indicates that the activated Na$_2$KSb(Cs) structure is effective not only in the visible but also in the near-infrared spectral range. These spectral changes are consistent with the evolution of the QE and photocurrent reported for polycrystalline Na$_2$KSb films and are typically attributed to the achievement of negative electron affinity \cite{holtom1979surface, interface}.

Fig. \ref{figpic4}(e) shows the LEED pattern observed on the surface of the graphene/6H-SiC(0001) substrate after vacuum annealing. The arrangement of the bright spots corresponds to the $(6\sqrt{3}\times6\sqrt{3})R30^\circ$ reconstruction, which is typical for single-layer and double-layer graphene on the SiC(0001) surface \cite{ni2008raman}. The presence of this LEED pattern indicates the atomic purity of the substrate surface after annealing. Importantly, a clear LEED pattern was observed after Na$_2$KSb growth, as shown in Fig. \ref{figpic4}(f). The pattern reveals two sets of spots with different intensities. These two sets are highlighted in blue and yellow in Fig. \ref{figpic4}(g). Given its FCC lattice, the unreconstructed Na$_2$KSb(111) surface exhibits a hexagonal Bravais lattice. The LEED pattern from such a surface would consist of only the blue or only the yellow reflections shown in Fig. \ref{figpic4}(g). Therefore, the observed pattern likely indicates the presence of two types of Na$_2$KSb(111) domains, rotated by 30° relative to each other. The intensity of the spots marked in blue is twice that of the yellow ones, suggesting a correspondingly higher surface coverage of the blue-oriented domains. Furthermore, the integral-order diffraction spots, such as (1,0) and (0,1), from the predominant Na$_2$KSb domain (shown in blue) are rotated by 15° relative to those of the graphene substrate, confirming the epitaxial nature of the growth at least in the initial stage. Fig. \ref{figpic4}(h) shows the LEED pattern measured after activation of the Na$_2$KSb surface. This pattern is similar to that observed on the Na$_2$KSb surface (Fig. \ref{figpic4}g). The persistence of this pattern indicates that the films retain their crystallinity after activation, suggesting no major changes occurred in the crystal lattice.

\begin{figure*}[t!]
\includegraphics[width=0.99\textwidth]{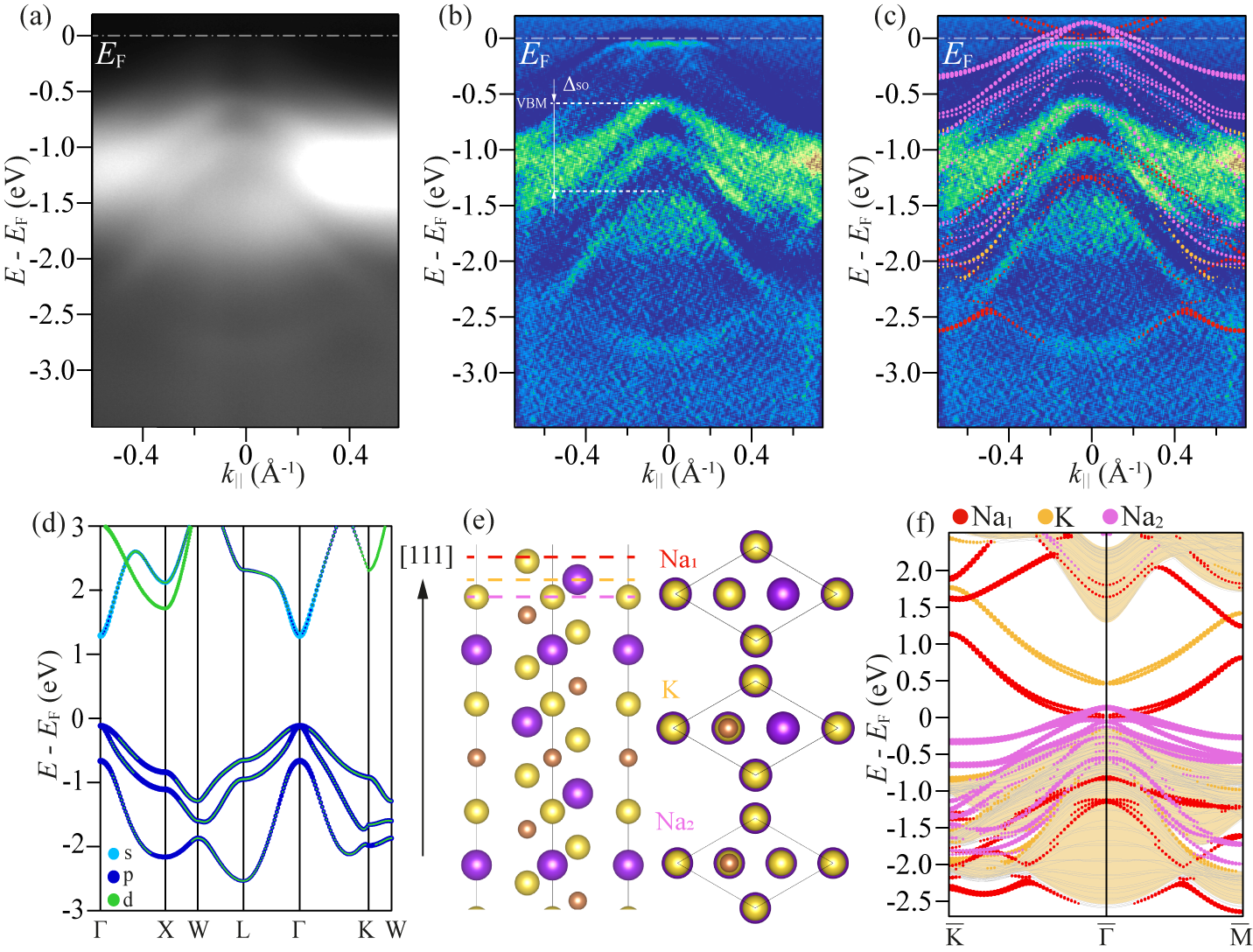}
\caption{\label{figpic}(a) ARPES spectrum of Na$_2$KSb measured at 77 K in  the $\bar{\mathrm K}-\bar\Gamma-\bar{\mathrm K}$ direction. (b) Smoothed second derivative of the spectrum shown in (a). (c) Direct comparison between the ARPES data from (b) and the calculated band structure (f). (d) Calculated bulk band structure of Na$_2$KSb. (e) Ball-and-stick model of the Na$_2$KSb(111) crystal structure. (f) DFT-calculated surface electronic structure of Na$_2$KSb(111).}

\end{figure*}

\subsection{\label{sec:level2}Electronic properties}

The crystalline ordered growth of Na$_2$KSb has enabled, for the first time, an experimental study of its electronic structure using ARPES, thereby allowing the validation of theoretical calculations. Fig. \ref{figpic}(a) shows the ARPES spectrum of the Na$_2$KSb(111) surface measured along the $\bar{\mathrm K}-\bar\Gamma-\bar{\mathrm K}$ direction of the predominant domain near the $\bar\Gamma$-point ($k = 0$ \AA$^{-1}$) at $T = 77$~K. Fig. \ref{figpic}(b) displays its smoothed second derivative (Laplacian), calculated using the curvature procedure in Igor Pro \cite{zhang2011precise}. The spectrum reveals a set of hole-like states and one electron-like state located 2.7 eV below the Fermi level ($E_\mathrm{F}$), with some of these states crossing $E_F$.  The obtained spin-orbit splitting, shown in Fig. \ref{figpic}(b), is 0.8 eV, which is slightly larger than the value predicted by DFT calculations for the bulk material (0.55 eV). The top of the valence band (VBM Fig. \ref{figpic}(b)) lies 0.58 eV below the $E_\mathrm{F}$.Given that the band gap of Na$_2$KSb is 1.4 eV \cite{interface}, the  Fermi level at the surface of the obtained photocathode is positioned near the middle of the gap, indicating that surface states pin the Fermi level at 0.58 eV above the visible bulk VBM.

Clearly, the experimentally measured electronic structure is more complex than the calculated bulk electronic structure of Na$_2$KSb typically described in the literature. Fig. \ref{figpic}(d) presents the results of DFT calculations for the bulk electronic states of Na$_2$KSb. Since ARPES is sensitive only to occupied electron states, only the p- and d-states forming the valence band are expected to be observed for non-degenerate Na$_2$KSb. However, these bulk states alone are insufficient to describe the electronic structure seen in the experiment, indicating that most of the features in Fig. \ref{figpic}(b) are of surface origin.

LEED measurements confirm that the grown sample has an (1x1) (111) surface. Three thermodynamically stable alkali terminations without reconstruction are possible: two sodium terminations (Na$_1$ and Na$_2$) and one potassium termination, indicated in Fig. \ref{figpic}(e) by differently colored dashed lines. Fig. \ref{figpic}(f) presents the DFT-calculated electronic structure of Na$_2$KSb(111) for each of these surface terminations. The corresponding colors denote the surface states associated with each isolated termination. To determine the contribution of surface states to the ARPES-measured electronic structure, the calculated surface states were overlaid on the Na$_2$KSb(111) spectrum (Fig. \ref{figpic}c). This overlay shows that describing all observed states requires the simultaneous presence of both sodium terminations on the surface. The presence of a potassium termination cannot be confirmed from the ARPES data, as its corresponding surface state lies mostly within the band gap, making it difficult to detect experimentally. Importantly, the calculations show that the band gap exhibits a high density of surface states originating from all terminations. And for all terminations, the emergence of surface states leads to a narrowing of the band gap, consistent with the behavior reported for alkali-terminated surfaces of cubic K$_2$CsSb \cite{wang2022first}. The high density of surface states further suggests that photoemission can occur not only via band-to-band transitions but also through direct excitation of electrons from surface states located below the Fermi level. According to our calculations Na$_1$ surface bands are composed from potassium $s$ and $p_z$, sodium $s$ and antimony $p_z$ orbitals, Na$_2$ surface bands are composed predominantly from antimony $p_x$ and $p_y$ orbitals, K terminated surface bands - from potassium $s$ and antimony $p_z$ and $s$ orbitals in the vicinity of $\bar\Gamma$-point. Thus, direct dipole excitation of an electron from surface band states to bulk conduction band is possible in all cases. Such transitions could play a significant role in the near-infrared region of the photocurrent spectral response, particularly near the absorption edge. Consequently, in this spectral region, the QE and even the spin polarization of the emitted electrons could vary for different crystal surfaces, such as Na$_2$KSb(111), (100), and (110), and should be systematically investigated in future studies.

\subsection{\label{sec:level2}Chemical composition}

\begin{figure}[h]
\includegraphics[width=0.45\textwidth]{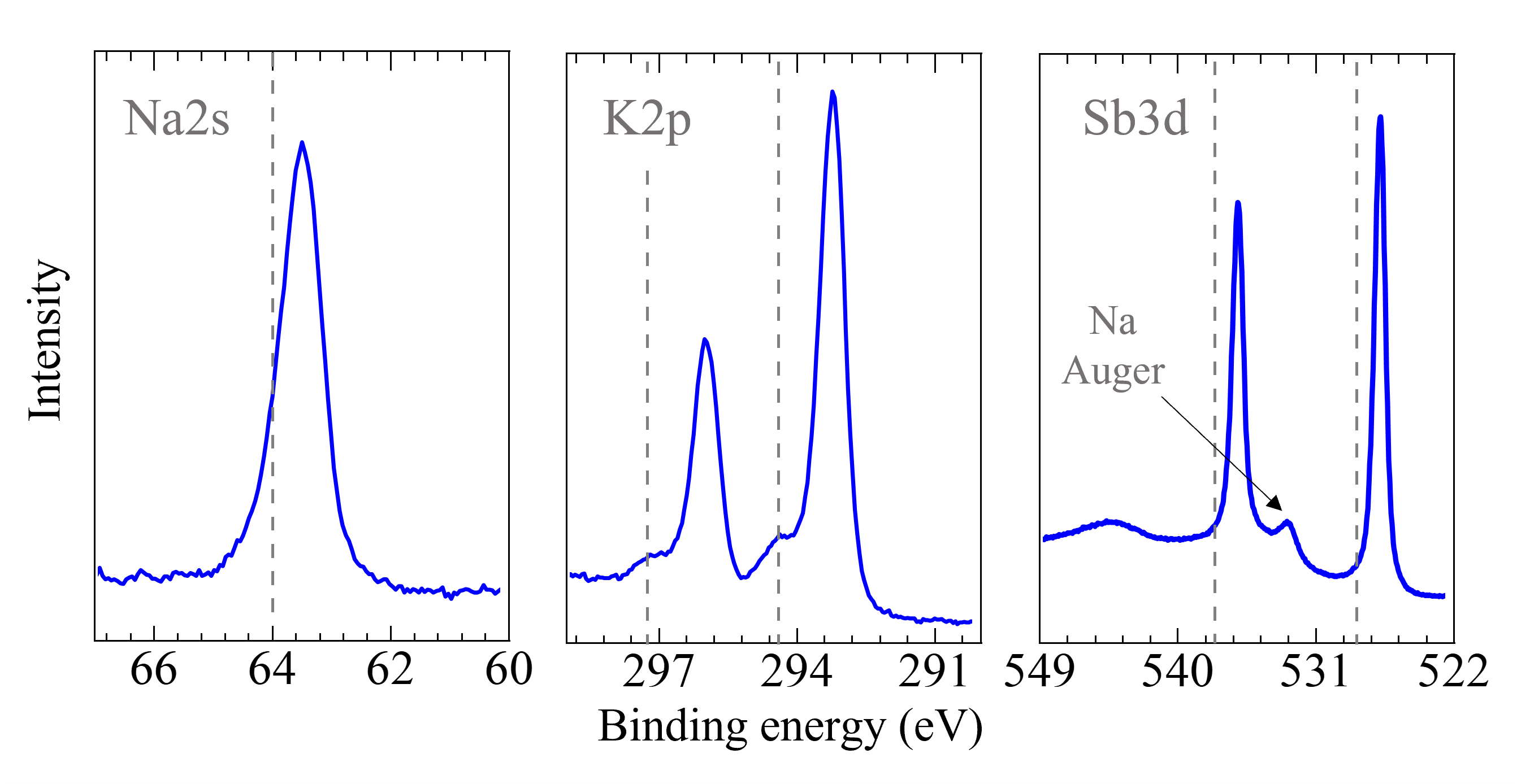}
\caption{\label{figxps}XPS spectra of the Na-$2s$, K-$2p$ and Sb-$3d$ core levels for Na$_2$KSb. The dashed lines indicate the positions of the elemental reference lines.}
\end{figure}

Since there are different stoichiometries of the multialkali antiminides of Na, K and Sb (Na$_2$KSb, NaK$_2$Sb, NaKSb), XPS measurements were performed to confirm the formation of Na$_2$KSb. The XPS spectra of the Na2s, K2p and Sb3d lines are shown in Fig. \ref{figxps}. The positions of the Sb3d XPS peaks confirm the formation of (Na,K)$_3$Sb. The Sb3d peaks are located at binding energies of 535.9 eV and 526.5 eV for the 3d5/2 and 3d3/2 spin-orbit components, respectively, consistent with the Sb$^{3-}$ state reported for various alkali antimonides \cite{parzyck2023atomically, panuganti2021synthesis, martini2015x}. The Na2s and K2p3/2 peaks are observed at 63.4 eV and 293.1 eV, respectively. Both peaks are shifted to lower binding energies relative to their elemental reference positions (indicated by gray dashed lines in Fig. \ref{figxps}), confirming chemical bond formation. Small shoulders are visible in the K $2p$ spectrum near the positions of the elemental lines, which may indicate a trace amount of metallic K on the surface. Contrary to the typical expectation — that alkali metals, due to their low electronegativity, would shift to higher binding energy upon charge transfer — all XPS core levels (Sb and alkali metals) in the present study exhibit shifts toward lower binding energy relative to their elemental positions. This effect has been observed and described elsewhere for various alkali antimonides \cite{martini2015x,schubert2013bi,parzyck2022single}. The stoichiometry was quantified from XPS data using SpecsLab Prodigy software. The APRES and LEED measurements presented in this paper were performed on two samples with stoichiometric compositions of Na52\% K15\% Sb33\% and Na45\% K22\% Sb33\%, which is close to Na$_2$KSb.

\FloatBarrier
\FloatBarrier

\section{\label{sec:level1}Conclusions}

In conclusion, we have successfully demonstrated the epitaxial growth of Na$_2$KSb films on graphene-coated SiC(0001) substrates via CVD. The high crystalline quality of the obtained surfaces enabled the first ARPES measurements of the Na$_2$KSb electronic structure, permitting direct validation of DFT calculations. These measurements revealed the presence of previously unreported surface states. Furthermore, the procedure of crystalline ordered growth opens the pathway for investigating the spin texture of electronic states using spin-resolved ARPES and its influence on the spin polarization of photoemitted electrons. Finally, we have shown that the crystalline order is preserved during activation with Cs and Sb, suggesting the potential for future ARPES studies to elucidate the influence of the activation process on the electronic structure and the mechanism of negative electron affinity formation.

\section{\label{sec:level1}Acknowledgments}

The authors acknowledge the support from the Russian Science Foundation (Grant No. 25-62-00004), SRF SKIF Boreskov Institute of Catalysis (FWUR-2024-0040) and ISP SB RAS.

\providecommand{\noopsort}[1]{}\providecommand{\singleletter}[1]{#1}%

\end{document}